\documentclass[preprint,aps,showpacs]{revtex4}

\usepackage{graphicx}
\usepackage{dcolumn}
\usepackage{epsfig}
\begin{document}
\title{Isotropic-nematic phase transition in suspensions of filamentous virus
and the neutral polymer Dextran}

\author{Zvonimir Dogic}\altaffiliation[current address: ]{Rowland Institute at Harvard, Cambridge, Massachusetts
02142}\affiliation{Complex Fluids Group, Physics Department,
Brandeis University, Waltham, Massachusetts 02454}

\author{Kirstin R. Purdy} \affiliation{Complex Fluids Group, Physics
Department, Brandeis University, Waltham, Massachusetts 02454}

\author {Eric Grelet}\altaffiliation[permanent address: ]{Centre de Recherche
Paul Pascal, CNRS UPR 8641, Pessac, France} \affiliation{Complex
Fluids Group, Physics Department, Brandeis University, Waltham,
Massachusetts 02454}

\author {Marie Adams}\affiliation{Complex Fluids Group,
Physics Department, Brandeis University, Waltham, Massachusetts
02454}

\author {Seth Fraden}\affiliation{Complex Fluids Group, Physics
Department, Brandeis University, Waltham, Massachusetts 02454}
\date{\today}

\begin{abstract}

We present an experimental study of the isotropic-nematic phase
transition in an aqueous mixture of charged semi-flexible rods
({\it fd} virus) and neutral polymer (Dextran). A complete phase
diagram is measured as a function of ionic strength and polymer
molecular weight. At high ionic strength we find that adding
polymer widens the isotropic-nematic coexistence region with
polymers preferentially partitioning into the isotropic phase,
while at low ionic strength the added polymer has no effect on the
phase transition. The nematic order parameter is determined from
birefringence measurements and is found to be independent of
polymer concentration (or equivalently the strength of
attraction). The experimental results are compared with the
existing theoretical predictions for the isotropic-nematic
transition in rods with attractive interactions.
\end{abstract}
 \maketitle

 \section{Introduction}

One of the fundamental notions of the theory of liquids, dating
back to van der Waals, is that liquid structure is determined by
the repulsive part of the intermolecular potential. The attractive
part of the potential determines the density of a liquid by
providing a cohesive background energy that is largely independent
of a particular configuration of
molecules\cite{Widom67,Gelbart80}. This is true as long as the
liquid is far from its critical point. Due to this reason there
has been a substantial effort over the past 50 years to use hard
spheres as a reference system to understand the behavior of all
simple liquids\cite{Hansen86}. Parallel to these endeavors, a
theory of a liquid of rods with purely repulsive anisotropic
interactions was developed by Onsager \cite{Onsager49}. It was
shown that this system exhibits an isotropic-nematic (I-N) phase
transition. The Onsager theory is based on the realization that
the virial expansion of the free energy converges for hard rods
with sufficiently large aspect ratio at the I-N phase transition,
in contrast to spheres where the virial expansion fails to
describe hard spheres at high concentration.

Once the behavior of a hard particle fluid is understood it is
possible to study the influence of attractions via a thermodynamic
perturbation theory~\cite{Gast86,Hansen86}. For hard spheres this
is relatively easy due to the fact that attractions provide an
structureless cohesive energy. In contrast, extension of the
highly successful Onsager theory valid for rods with short range
repulsions to a system of rods with longer range attractive
interactions is much more difficult. The difficulties stem from
the fact that attractive rods are in their lowest energy state
when they are parallel to each other. These are exactly the
configurations that need to be avoided if the second virial term
on which the Onsager theory is based, is to accurately describe
the system~\cite{Schoot92}. In one study the Onsager functional
has been straightforwardly extended to include an additional
attractive interaction\cite{Warren94} (from now on called Second
Virial Theory with Attraction (SVTA)). Because of the problems
already mentioned the author argues that SVTA is valid only for
very weak attractions. Indeed, at high strengths of attraction
un-physical states such as a collapse to infinitely dense state
are predicted. The physical picture that emerges from the SVTA
theory is that of a van der Waals like liquid of rods where its
primary structure (i.e. nematic order parameter) is determined
purely by the repulsive interactions, while attractions serve as a
uniform structureless glue holding the rods together at a given
density.

For the theory to work at all densities the free energy of the
unperturbed liquid of rods needs to take into account third and
higher virial coefficients. An alternative theory that
accomplished this uses scaled particle free energy of hard rods as
a basis to study the influence of attractive interactions on the
I-N phase transition~\cite{Lekkerkerker94,Bolhuis97}(from now on
called Scaled Particle Theory with Attractions (SPTA)). The scaled
particle expression for hard rods includes third and higher virial
coefficients. Therefore it is reasonable to expect that this
theory would be more accurate at higher rod and/or polymer
concentrations. An additional advantage of the SPTA theory is that
it does not assume that the depletion interaction is pairwise
additive. Computer simulations have shown that pairwise additivity
of the intermolecular potential assumption is not an adequate
approximation when the radius of the polymer is larger then the
radius of the colloid~\cite{Meijer94,Gelbart80}.

In this paper we experimentally study the influence of attractive
interactions on the I-N transition and compare them to both SVTA
and SPTA theory. As a reference system we use an aqueous
suspension of semi-flexible rod-like {\it fd} viruses. Previous
work has shown that the behavior of {\it fd} virus is consistent
with the theoretical predictions for semi-flexible rods with
purely repulsive interactions
~\cite{Tang95,Fraden95,Dogic97,Dogic00c}. Strictly speaking, {\it
fd} forms a cholesteric phase and undergoes an
isotropic-cholesteric transition, but because the free energy
difference between a cholesteric and a nematic phase is small we
refer to the cholesteric phase as nematic in this paper.
Additionally, introducing a finite flexibility into hard rods
significantly alters both the location and nature of the
isotropic-nematic and nematic-smectic phase
transition\cite{Khokhlov81,Dogic97}. Here we show that flexibility
also changes the isotropic-nematic phase transition in rods with
attractive interactions.

We induce attraction experimentally by adding a non-adsorbing
polymer to the colloidal suspensions, which leads to the depletion
interaction where the range and the strength of the attractive
potential is controlled by the polymer size and concentration
respectively~\cite{Asakura58}. Although this work specifically
deals with a colloid/polymer mixture its results are of a general
significance to other anisotropic fluids which have attractive
interactions due to other reasons (i.e van der Waals attractions).
The main difference between polymer induced depletion attractions
and attractions due to van der Waals forces is that in the
depletion case there is partitioning of the polymer between
coexisting phases \cite{Lekkerkerker92a}. Therefore the strength
of the interaction between two colloids depends on the phase in
which the colloids are located.

There have been previous experiments on the influence of polymer
on the I-N phase transition in mixtures of boehmite rods and
polystyrene polymers and mixtures of cellulose nanocrystals and
Dextran polymers~\cite{Buitenhuis95,Bruggen00,Edgar02}. Other
studies related to our work have focused on the condensation of
rod-like polymers due to the presence of polymer and/or
multivalent cations~\cite{Tang96a,Vries01}. The conditions in
those studies correspond to the upper left corner of the phase
diagram in Fig.~\ref{DEX500}. We also note that at very high
polymer concentrations the fd system exhibits a direct
isotropic-smectic coexistence and a number of metastable complex
structures associated with this transition have been described
elsewhere\cite{Dogic01,Dogic03}.

In this paper we limit ourselves to the I-N transition. In Section
II we present the experimental details of our measurements. In
section III the effective intermolecular potential acting between
two rod-like particles is discussed. In section IV we present the
measured phase diagrams as a function of ionic strength and
polymer size, and in section V we present our conclusions. In the
appendix we provide the formulas necessary to calculate the phase
diagrams in the SPTA theory.

\section{Materials and Methods}

\begin{figure}
\centerline{\epsfig{file=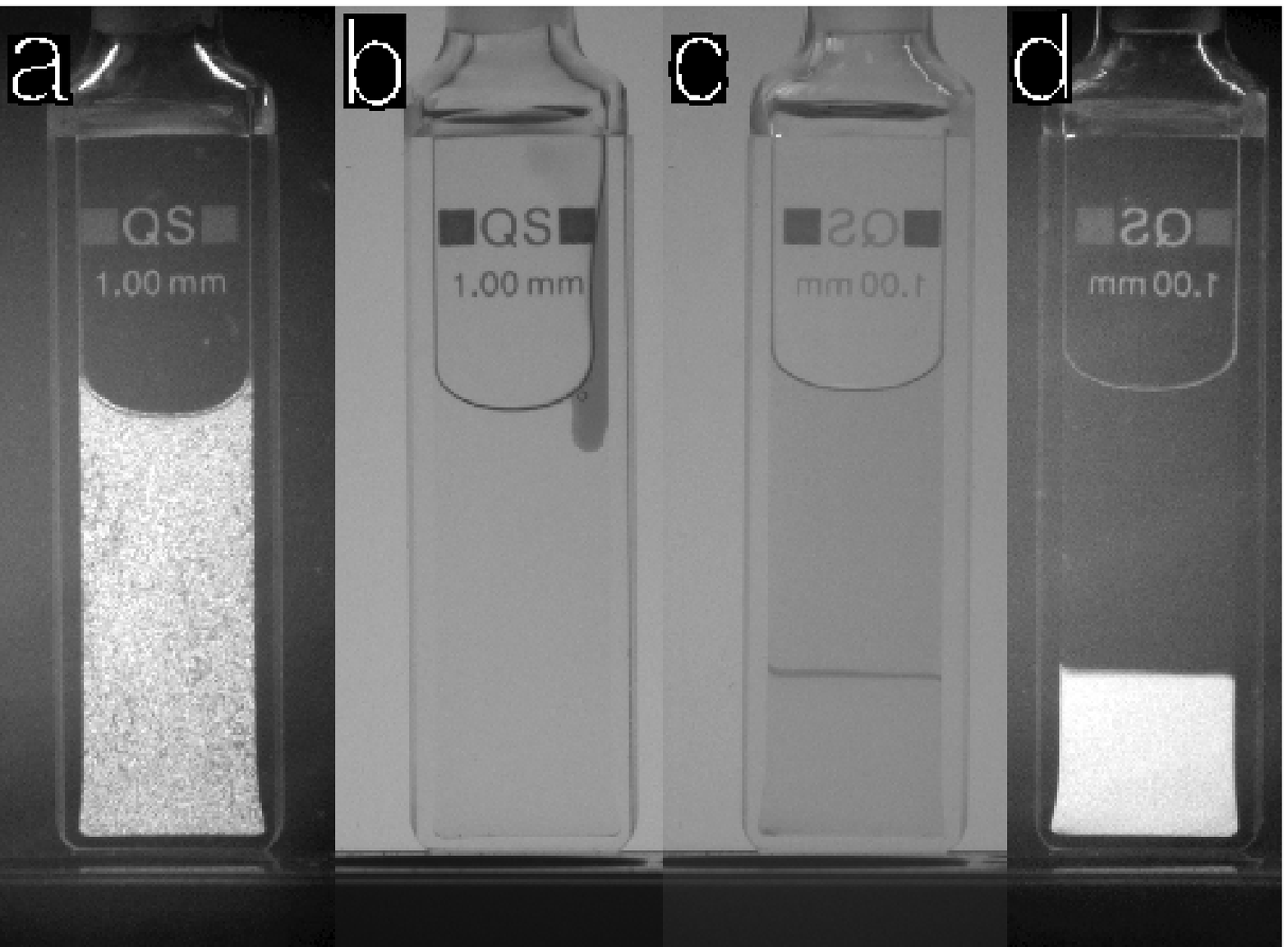,width=8.cm}}
\caption{\label{cartoon}(Color online) A sequence of images
illustrating the preparation of a sample which was used in
determining the phase diagram. a) A nematic liquid crystal of {\it
fd} virus in buffer between crossed polarizers showing disordered
birefringent domains. b) A highly concentrated solution of Dextran
labelled with yellow fluorescein is added to the transparent {\it
fd} nematic liquid crystal. c) After the sample in (b) is
vigorously shaken it phase separates into the coexisting nematic
and isotropic phases. The macroscopic phase separation takes from
few hours to couple of days depending on the location in the phase
diagram. d) Same sample as image (c) but taken between crossed
polarizers. Image shows dense birefringent nematic phase on the
bottom and Dextran rich isotropic phase on the top which is yellow
in appearance. Images (a) to (d) are taken on the same sample. }
\end{figure}

Bacteriophage {\it fd} was grown and purified as described
elsewhere~\cite{Maniatis89}. $\it Fd$ is a rod-like semiflexible
charged polymer of length 0.88 $\mu$m, diameter 6.6 nm,
persistence length 2.2$\mu$m and surface charge density of
1e$^-$/nm at pH 8.15. All samples where dialyzed against 20 mM
Tris buffer at pH=8.15 and NaCl was added until the desired ionic
strength was achieved. Dextran and FITC-Dextran with molecular
weights (MW) of 500,000 and 150,000 g/mol (Sigma, St. Louis, MO)
were used as the non-absorbing polymer and dissolved in the same
buffer solution. The samples are prepared in the two phase region
of the phase diagram as is shown in Fig.~\ref{cartoon}.
Concentrations of coexisting phases were measured using absorption
spectrophotometry. The optical density of {\it fd} is
$\mbox{OD}_{\mbox{\scriptsize{269 nm}}}^{\mbox{\scriptsize{1
mg/ml}}}=3.84$ for a path length of 1 cm. To determine the
concentration of Dextran polymer we used a mixture of 95\% Dextran
and 5\% FITC labelled Dextran. The optical density of FITC-Dextran
was determined by dissolving a known amount of polymer in a buffer
solution and measuring the OD at 495nm. The relationship between
the radius of gyration $R_g$ of Dextran and its molecular weight
(MW) in units of g/mol is $R_g[\mbox{\AA}]=0.66
(\mbox{MW})^{0.43}$ ~\cite{Senti55}. The reason for the small
exponent 0.43 is due to the fact that Dextran is a branched
polymer. The volume fraction of polymer $\phi_{\mbox{\scriptsize
polymer}}$ was calculated by $\phi_{\mbox{\scriptsize
polymer}}=\rho \frac{4}{3} \pi R_g^3$, where $\rho$ is the polymer
number density. The order parameter of the nematic phase was
measured with a Berek compensator, by placing the suspension into
a quartz x-ray capillary with a diameter 0.7 mm (Charles Supper,
Natick, MA). Samples were aligned with a 2T magnetic
field~\cite{Oldenbourg86} and the birefringence was measured. The
order parameter ($S$) is obtained using the relationship $\Delta
n= S \rho_{\mbox{\scriptsize fd}} n_o $ where
$\rho_{\mbox{\scriptsize fd}}$ is the number of rods per unit
volume of {\it fd} virus, $\Delta n$ is the birefringence measured
using Berek compensator on an Olympus microscope and S is the
nematic order which varies between 0 for the isotropic phase and 1
for a perfectly aligned phase. The birefringence of perfectly
aligned {\it fd}, $n_0=3.8\times10^{-5}\pm0.3\times10^{-5}$ ml/mg,
was recently obtained from x-ray experiments\cite{Purdy03}.

\section{Intermolecular potential}

When a colloid is suspended in a polymer solution it creates
around itself a shell from which the center of mass of a polymer
is excluded. When two colloids approach each other there is an
overlap of the excluded volume shells which leads to an imbalance
of the osmotic pressure that is exerted on each colloid. This
results in an effective attractive potential known as the
depletion potential. In the Asakura-Oosawa model (AO), polymers
are assumed to behave as spheres of radius
$R^{\mbox{\scriptsize{AO}}}$, which can freely interpenetrate each
other but interact with colloids via hard core repulsive
interactions~\cite{Asakura58}. The relationship between
$R^{\mbox{\scriptsize{AO}}}$ and radius of gyration of a polymer
($R_g$) is $R^{\mbox{\scriptsize{AO}}}=2R_g/\sqrt\pi$. This
approximation is valid as long as the size of the colloidal
particle is much larger then the radius of the penetrable sphere
$R^{\mbox{\scriptsize{AO}}}$~\cite{Chatterjee98,
Eisenriegler99,Tuinier00}. If the size of a colloid is equal to or
smaller then $R^{\mbox{\scriptsize{AO}}}$, the colloid can
penetrate into the open polymer structure without overlapping any
of the polymer segments. In this case the range and the depth of
the attractive depletion potential will be significantly weaker
when compared to the predictions of the AO model. In our
experiments the diameter of the polymer is up to 5 times the
diameter of the rod-like virus and therefore we expect that the
depletion potential significantly deviates from the Asakura-Oosawa
penetrable sphere model.

Since there are no analytical results on the depletion potential
between rod-like colloids we estimated it using computer
simulations. The method used to obtain the potential is described
in detail in the paper by Tuinier {\it et. al.}\cite{Tuinier00};
here we briefly outline the procedure. Two spheres, cylinders or
walls are set at a fixed distance apart and an attempt is made to
insert a non-self-avoiding polymer molecule at random positions.
When simulating the depletion potential between the cylinders they
are oriented in perpendicular directions. If any segment of the
polymer overlaps with either colloid, the insertion attempt fails
and the polymer is not counted. The profile of the depletion
potential is then equal to

\begin{equation}
U_{\mbox{\scriptsize depletion}}(x)=k_BT (N(\infty)-N(x))
\end{equation}

\noindent where $N(x)$ is the number of polymers successfully
inserted in the simulation box when two colloidal objects are a
distance $x$ apart. $N(\infty)$ is the number of polymers inserted
when two colloids are apart at a distance which is much larger
then the range of the intermolecular potential.

\begin{figure}
\centerline{\epsfig{file=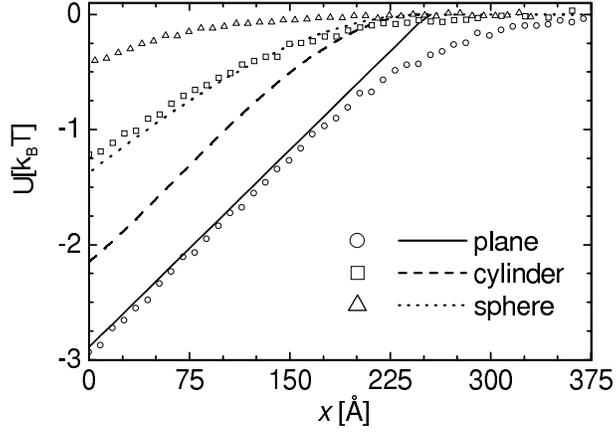,width=8cm}}
\caption{\label{Depletion_potential} Depletion potential $U$
between two walls, two perpendicular cylinders and two spheres
obtained from computer simulation are shown by open spheres,
squares and triangles respectively. In the two wall simulation the
wall size was $313 \times 313\mbox{\AA}^2$ and periodic boundary
conditions where used. The diameter of the spheres and cylinders
is 66 $\mbox{\AA}$ while $R_g$ of the polymer is 111 $\mbox{\AA}$.
The lines indicate depletion potentials as predicted by the
penetrable sphere (AO) model. The separation $x$ is the closest
distance between two surfaces.  The number concentration of the
polymer $\rho$ is equal to the overlap concentration $\rho=3/(4\pi
R_g^3)$, while the radius of the penetrable spheres is
$R^{\mbox{\scriptsize AO}}=2 R_g/\sqrt{\pi}=125$ $\mbox{\AA}$. The
AO theory overestimates the potential between spheres and between
cylinders because polymer deform around colloidal particles.}
\end{figure}

The depletion potentials between walls, spheres and rods obtained
from the simulations are shown in Fig.~\ref{Depletion_potential}.
From the exact results, it is known that the depletion potential
at small separations between two parallel walls induced by AO
penetrable spheres is equivalent to the depletion potential
induced by polymer (without excluded volume interactions), if
$R^{\mbox{\scriptsize AO}}=2R_g/\sqrt\pi$
~\cite{Asakura54,Tuinier00}. If we use this fact, the simulation
results for the depletion potential between two plates (indicated
by open circles in Fig.~\ref{Depletion_potential}) are in a very
good agreement with the potential predicted by the AO theory
(indicated by the full line in Fig.~\ref{Depletion_potential}), as
long as the separation between the plates is smaller then $3
R_g/2$. At larger separations we observe that the potential
exerted by the polymer has longer range attraction than the
equivalent penetrable sphere, as was previously
noted~\cite{Tuinier00}. This is because a polymer is only
spherical on average and will adopt elongated conformations on
occasion. The simulation results for the depletion potential
between two spheres immersed in a polymer suspension with
$R_g/R_{\mbox{\scriptsize colloid}}=3.36$ is significantly weaker
then what is predicted by the penetrable AO sphere model. The
reason for this is that a small sphere has a high probability of
penetrating a polymer with a large radius of gyration since
polymers have very open structures. The rods have a profile of an
infinite plane in one direction and a profile of a sphere in the
other direction. It follows that a cylinder with the same diameter
as a sphere is less likely to interpenetrate with a polymer coil.
Therefore the depletion interaction between cylinders is stronger
than between spheres of equal diameter and weaker then the
depletion interactions between two walls. Even for the case of
cylinders, the potential obtained from the AO model significantly
overestimates the strength of the potential obtained from the
simulation as is shown in Fig.~\ref{Depletion_potential}. In this
paper we assume that the strength of the depletion potential
between two cylinders oriented at an angle $\gamma$ scales as $1/
\sin \gamma$, but the shape remains independent of $\gamma$. To
verify this we have simulated the potential between two cylinders
that are either parallel or perpendicular to each other. For these
two cases we obtain depletion potentials that are almost identical
to each other after they are rescaled by a constant. This supports
our assumption that the shape of the depletion potential is
independent of the cylinder orientation.

\begin{figure}
\centerline{\epsfig{file=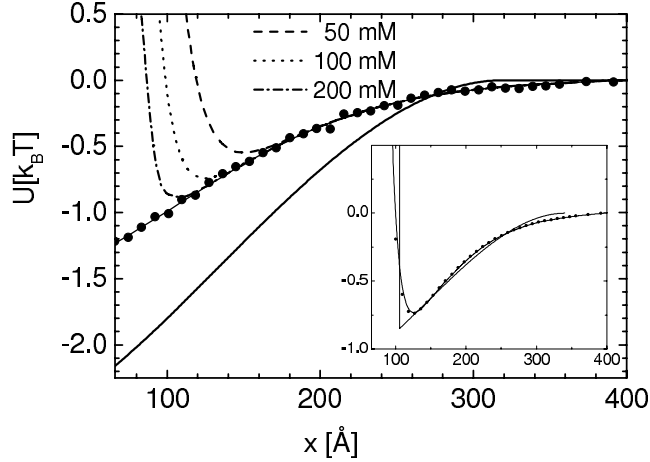,width=8.2cm}} \caption
{\label{potential} Total interaction potential $U$ as a function
of separation $x$ between two viruses ($D=66$ $\mbox{\AA}$)
oriented at $90^{\circ}$ with respect to each other and immersed
in a suspension of polymers of concentration $\rho=3/(4\pi R_g^3)$
and radius $R_g=111$ $\mbox{\AA}$ at three different ionic
strengths. The interaction potential is a sum of electrostatic
repulsion and depletion interaction. The effect of electrostatic
repulsion for {\it fd} with net linear charge density 1 $e^-$/\AA
of is accounted for by treating the the {\it fd} as a hard
particle with a larger effective diameter $D_{\mbox{\scriptsize
eff}}$\cite{Tang95,Vroege92}. Filled circles indicate the
depletion potential obtained from Monte Carlo simulation of
polymers without excluded volume interactions. Since the polymer
diameter is larger then the rod diameter, the polymer and rod-like
viruses can easily interpenetrate. This results in a effective
depletion attraction which is smaller than what is predicted by
the AO model (indicated by the full line). The phase diagrams
corresponding to these interaction potentials are shown in
Fig.~\ref{DEX150_phase}. Inset : In theoretical calculations we
approximate the intermolecular potential between rods with an
effective hard core diameter $D_{\mbox{\scriptsize
eff}}$\cite{Tang95} and attractive potential. The attractive part
of the potential is modeled by AO penetrable spheres whose
effective radius and concentration best fits the potential
obtained through computer simulation. This effective
intermolecular potential is compared to the potential obtained
through the computer simulation in the inset. In the inset
$\rho_{\mbox{\scriptsize eff}}/\rho=0.39$ and
$R^{\mbox{\scriptsize AO}}_{\mbox{\scriptsize
eff}}/R^{\mbox{\scriptsize AO}}=0.99$.}
\end{figure}

The total interaction potential between two {\it fd} viruses in a
{\it fd}/Dextran mixture is a combination of hard core repulsion,
a steep short range electrostatic repulsion, and the longer range
depletion attraction described above. As the ionic strength
decreases both the range and the depth of the potential decreases
as is shown in Fig.~\ref{potential}. This is due to the fact that
Dextran is an uncharged polymer and therefore the depletion
attraction is independent of the ionic strength. Decreasing the
ionic strength results in longer range electrostatic repulsion
which screens out ionic strength independent depletion attraction.

The short range electrostatic repulsion and longer range depletion
attraction scale as $1/\sin(\gamma)$ where $\gamma$ is the angle
between two rods. Therefore, to a first approximation the position
of the minimum of the intermolecular potential does not change
when the angle between two rods changes; only the magnitude of the
minimum changes. To account for the rapidly decaying electrostatic
repulsion we re-scale the hard core diameter to an effective hard
core diameter $D_{\mbox{\scriptsize eff}}$ as was described
previously~\cite{Tang95}. We note however that the use of
$D_{\mbox{\scriptsize eff}}$ is rigourously justified only in the
dilute regime where the second virial coefficient quantitatively
describes the system, ie. at the isotropic-nematic transition of
pure rod suspensions. Therefore one of the causes of the
discrepancy between theory and experiments stems from our crude
treatment of the electrostatic interactions. As discussed
previously the attractive depletion potential is weaker then the
predictions of the AO model. To account for this in the
calculation of the phase diagram we simulated the depletion
potential for experimentally relevant parameters. The simulated
potential is mapped onto AO model where the effective
concentrations of the interpenetrable spheres
($\rho_{\mbox{\scriptsize eff}}$) and effective polymer radius
($R_{\mbox{\scriptsize eff}}^{\mbox{\scriptsize AO}}$) are
adjusted to yield the best fit to the simulated potential. We
define $\rho_{\mbox{\scriptsize eff}}=\alpha \rho$ where
$\rho=N/V$ is the actual number density of AO penetrable hard
spheres and $R^{\mbox{\scriptsize AO}}_{\mbox{\scriptsize
eff}}=\beta R^ {\mbox{\scriptsize AO}}$. Surprisingly we find that
the $\alpha$ is much smaller then 1 while $\beta$ is only slightly
smaller than 1 for parameters used in our experiments. This can be
seen in Fig.~\ref{Depletion_potential} where the range of the
depletion potential between two spherocylinders for the AO
penetrable sphere model is almost identical to the range of the
simulated potential while the depth is very different. The reason
for this is that the AO model underestimates the depth of the
depletion attraction at large distances
(Fig.~\ref{Depletion_potential}). If the polymer size is increased
further we observe that the value of $\beta$ will start decreasing
rapidly. The comparison between the simulated potential and the
effective potential used in the theoretical calculations of the
phase diagrams is shown in the inset of Fig.~\ref{potential}. The
phase diagrams are calculated using the effective rod diameter
$D_{\mbox{\scriptsize{eff}}}$, the effective polymer radius
$R^{\mbox{\scriptsize AO}}_{\mbox{\scriptsize eff}}$, and the
effective polymer concentration $\rho_{\mbox{\scriptsize eff}}$.
The calculation of the SPTA and SVTA phase diagrams for
semi-flexible rods is described in the Appendix. Once the phase
diagrams are obtained the polymer concentrations are rescaled back
to the actual volume fraction of polymer. Specifically we
calculate the theoretical phase diagrams using
$\rho_{\mbox{\scriptsize eff}}$ and then in order to compare with
experiment we plot the theoretical results using
$\rho=\rho_{\mbox{\scriptsize eff}}/\alpha$.

\section{Results}

\begin{figure}
\centerline{\epsfig{file=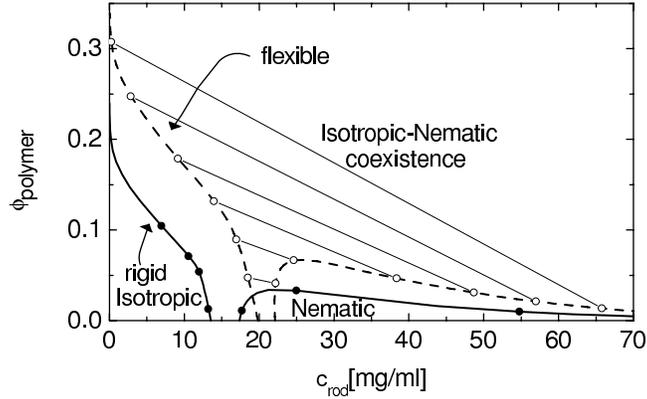,width=8.5cm}}
\caption{\label{flexibility} Phase diagram for rigid and
semi-flexible rods calculated using the SPTA theory. The boundary
between the isotropic(I)-nematic(N) two phase region and the
region where a single phase is stable is indicated by the thick
dashed line for semi-flexible rods and thick full lines for rigid
rods. Tie lines between the coexisting phases are shown by thin
lines. For the flexible particle the ratio of the contour length
to persistence length is $L/P=0.4$. The phase diagram was
calculated using $\delta=84$ and $q=2.2$. The polymer
concentration is defined as follows $\phi_{\mbox{\scriptsize
polymer}}=\rho\frac{4 \pi R_g^3}{3}$.}
\end{figure}

\begin{figure}
\centerline{\epsfig{file=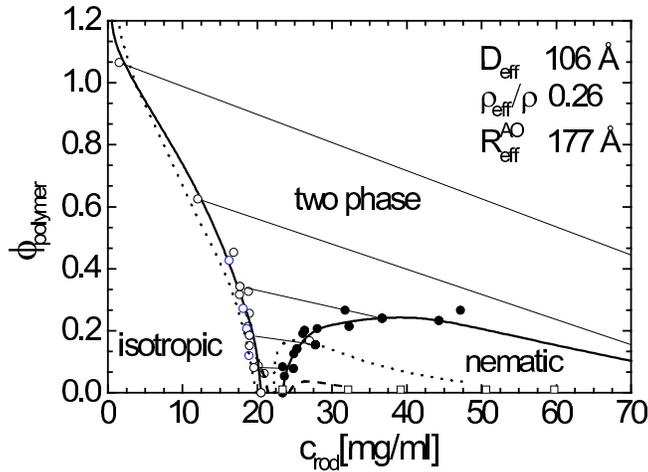,width=8.5cm}}
\caption{\label{DEX500} Phase diagram for a mixture of {\it fd}
virus and Dextran (MW 500,000, $R_g=176$ $\mbox{\AA}$ or
$R^{\mbox{\scriptsize AO}}=199$ $\mbox{\AA}$) at 100 mM ionic
strength. The measured points indicate the rod and polymer
concentrations of the coexisting isotropic and nematic phases. The
full line is a guide to the eye indicating the two phase region.
Tie-lines are indicated by thin full lines. The SPTA and SVTA
predictions are indicated by the dotted lines and dashed lines
respectively.}
\end{figure}

In Fig.~\ref{flexibility} a typical phase diagram for a mixture of
hard rigid rods and polymers is indicated by thick full lines. As
was shown in previous work by Lekkerkerker et. al.
\cite{Lekkerkerker94} adding polymer widens the isotropic-nematic
coexistence and leads to partitioning of the polymer between
isotropic and nematic phases. In the same figure dashed lines
indicate the phase diagram of a mixture of semi-flexible rods and
polymers. The influence of the flexibility on the
isotropic-nematic phase transition is well studied for the case of
rods with hard core repulsive interactions~\cite{Vroege92,Chen93}.
Flexibility increases the concentration of the I-N co-existence,
decreases the width of the I-N coexistence, and reduces the order
parameter of the nematic phase co-existing with the isotropic
phase. In Fig.~\ref{flexibility} the theoretical phase diagrams
for two equivalent systems of rods with attractions are shown with
the only difference being the flexibility of the rod. For the case
of the rigid rods the concentration of the polymer needed to
induce widening of I-N coexisting phases is much lower than for
that of semi-flexible rods. This is due to the fact that to
compress semi-flexible rods, the polymer has to work against both
rotational and internal bending contributions to the entropy. We
conclude that flexibility also suppresses the formation of the
nematic phase in attractive rods. Next we proceed to compare the
theoretical phase diagrams to experiments.

A representative experimental phase diagram of an {\it fd}/Dextran
mixture at high ionic strength is shown in Fig.~\ref{DEX500}. Two
features of the phase diagram are in qualitative agreement with
the theoretically predicted one. First, introducing attractions
widens the isotropic-nematic coexistence. Second, at intermediate
polymer concentrations polymer preferential partitions into the
isotropic phase. At very high polymer concentration the rods and
polymers are essentially immiscible with a nematic phase of pure
rods coexisting with an isotropic phase of pure polymers. This
part of the phase diagram has been measured elsewhere
\cite{Dogic03}.

\begin{figure}
\centerline{\epsfig{file=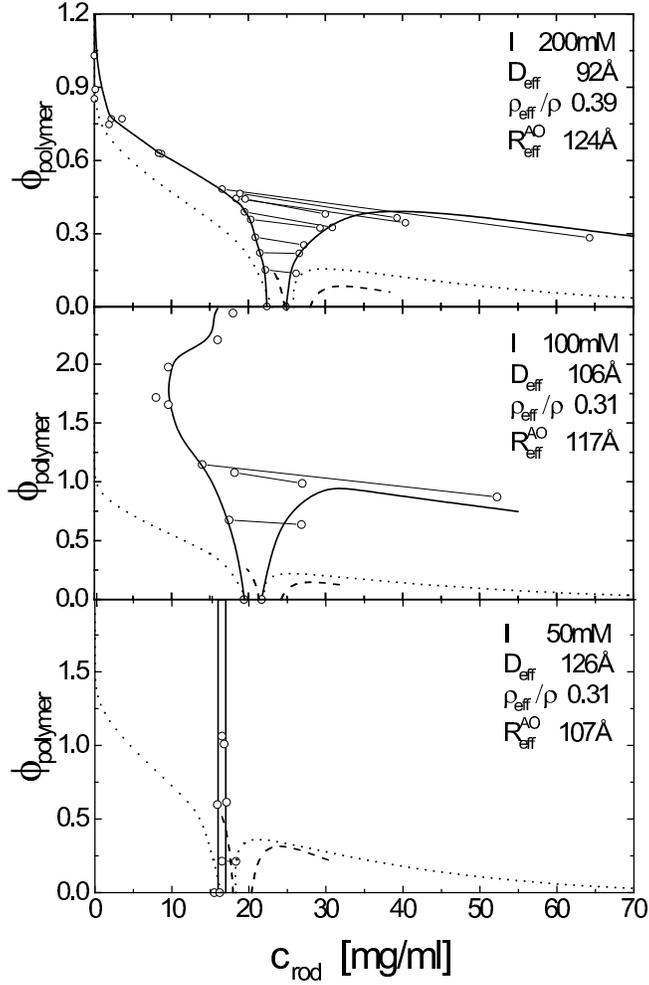,width=8.5cm}}
\caption{\label{DEX150_phase} Phase diagrams of a mixture of {\it
fd} virus and Dextran polymer (MW 150,000, $R_g=111$ $\mbox{\AA}$
or $R^{\mbox{\scriptsize AO}}=125$ $\mbox{\AA}$) at 50 mM, 100 mM
and 200 mM ionic strength. Coexisting phases are indicated by open
circles while the full line is an eye guide separating two phase
region from isotropic and nematic phases. The predictions of the
SPTA and SVTA theories are indicated with dashed and dotted lines
respectively. The polymer concentration is defined as follows
$\phi_{\mbox{\scriptsize polymer}}=\rho\frac{4 \pi R_g^3}{3}$.}
\end{figure}

We proceed to study the influence of the ionic strength on the
phase behavior. Changing ionic strength significantly modifies the
interaction potential as was shown in Fig.~\ref{potential}. The
phase diagrams at three different ionic strength are shown in
Fig.~\ref{DEX150_phase}. The experimentally measured phase diagram
at 50 mM ionic shows that the addition of the polymer has no
effect on the coexistence concentrations of the I-N transition.
This is in stark disagreement with theory which predicts strong
partitioning of the polymer. The implication from these
experimental results is that the depletion attraction is
completely screened by the long range electrostatic repulsion. In
calculating the potential energy between charged rods in the
presence of neutral polymer we are summing two large terms of
opposite signs (Fig. \ref{potential}). Small inaccuracies in the
theory of either of these terms could account for the discrepancy
between the theoretical and experimentally observed phase
diagrams.

As the ionic strength is increased to 100 mM the addition of the
polymer initially increases the width of the co-existence
concentration, while at very high polymer concentration we observe
re-stabilization of the I-N transition. This was also observed in
mixtures of {\it fd} and Dextran (MW 500,000) at 100 mM. This
observation can be explained by the fact that restabilization of
the I-N phase transitions occurs when the polymer is in the
semi-dilute regime. In this regime the range of the depletion
interaction is of the order of the correlation length (polymer
mesh size), which is smaller then the radius of
gyration~\cite{Verma00}. Moreover the correlation length decreases
with increasing concentration. Since the range of attraction
decreases in the semi-dilute regime, the long range electrostatic
repulsion will screen out any depletion attraction in the
semi-dilute regime.  This mechanism of depletion re-stabilization
was previously observed in mixtures of charged spherical colloids
and polymer mixtures in aqueous suspension \cite{Gast83}.

At the highest ionic strength of 200 mM a relatively low
concentration of polymer is needed to induce a complete phase
separation between a polymer-rich, rod-poor isotropic phase and a
rod-rich, polymer-poor nematic phase. At this ionic strength no
reentrant I-N phase behavior is observed for all accessible
polymer concentrations. The phase behavior at this ionic strength
has the same qualitative behavior as predicted by both SVTA and
SPTA. However, when the theory is quantitatively compared to
experiment there are large differences between the predicted phase
boundaries (full lines) and experimentally measured phase
boundaries. Perhaps the fact that the disagreement between theory
and experiment is worst at low ionic strength and high rod
concentration indicates that our approximation of treating
electrostatically repulsive rods as hard rods with an effective
diameter $D_{\mbox{\scriptsize eff}}$ is invalid under these
conditions, as previously mentioned.

\begin{figure}
\centerline{\epsfig{file=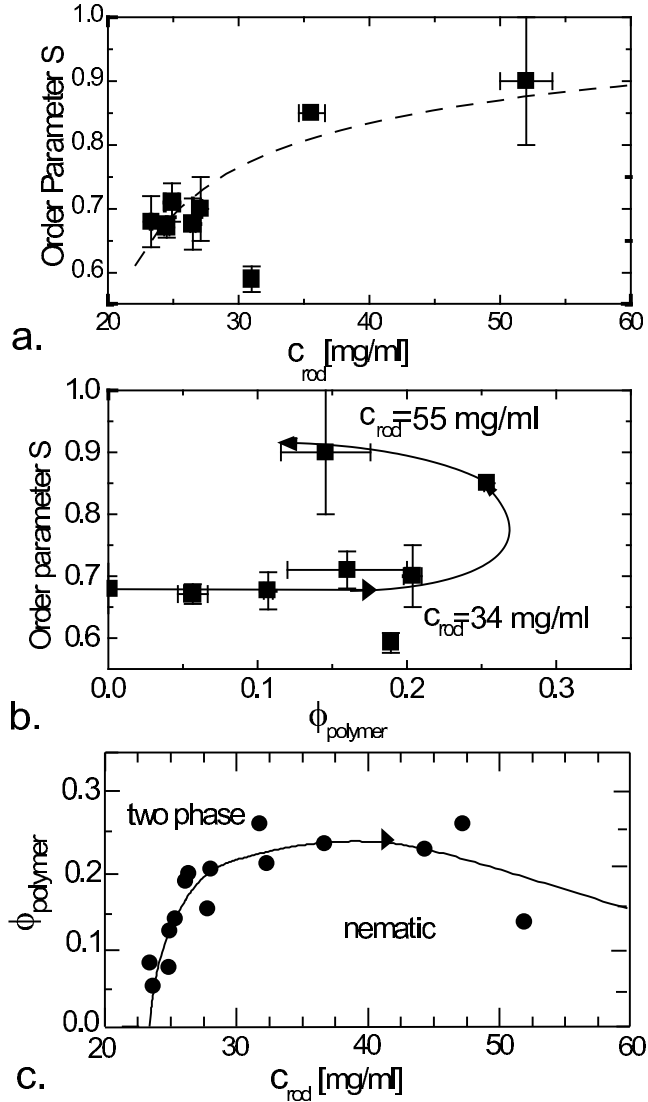,width=8.5cm}}
\caption{\label{SCphi} Measurements of the order parameter of the
nematic phase in coexistence with the isotropic phase for a
mixture of {\it fd} (rod) and Dextran (M.W. 500,000, polymer) at
100 mM ionic strength. Order parameter (S) is graphed as a
function of {\it fd} concentration (a) and Dextran concentration
(b) for the coexisting nematic concentrations shown in (c). The
order parameter is double valued in (b) because along the nematic
branch of the coexistence curve there are two different rod
concentrations with the same polymer concentration as shown in
(c). Data in (c) is the same as that shown in Fig. \ref{DEX500}.
Dashed line in (a) indicates the theoretical dependence of the
order parameter on the concentration of rods as obtained using
scaled particle theory. This relationship agrees well with
experimental data for {\it fd} at high ionic strength using x-ray
scattering~\cite{Purdy03}. Arrows in (b) and (c) indicate the
direction of increasing {\it fd} concentration. Below
$\phi_{\mbox{\scriptsize{polymer}}}\sim 0.2$ the nematic {\it fd}
concentration is essentially constant.}
\end{figure}

\begin{figure}
\centerline{\epsfig{file=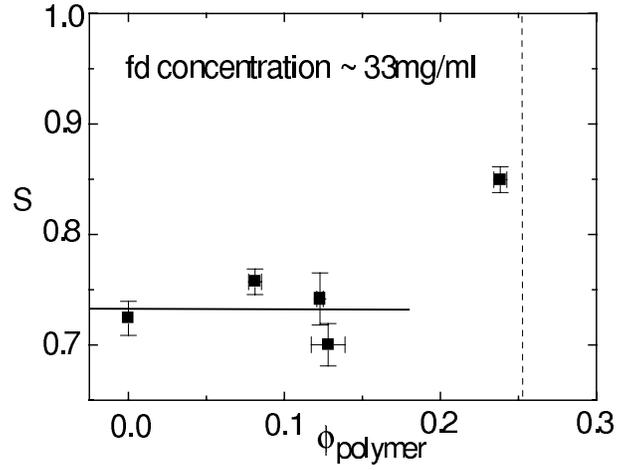,width=8cm}}
\caption{\label{9.fig} The order parameter of the nematic phase of
the {\it fd}/Dextran (M.W. 500 000) mixture at 33 mg/ml {\it fd}
and 100 mM ionic strength as a function of increasing polymer
concentration. The horizontal line drawn is a guide to the eye
showing the independence of the nematic order parameter with
polymer concentration. The vertical line indicates the location of
the nematic-isotropic transition. }
\end{figure}

If attractions are introduced to a hard sphere system, the
assembly will decrease its energy by decreasing the average
separation between spheres, which in turn increases the density of
the stable liquid phase. Unlike spheres, rod-like particles with
attraction have a more complex interaction potential. They can
lower their interaction energy not only by decreasing their
separation (increasing their density), but also by decreasing
their relative angle (increasing their order parameter), or a
combination of both. To distinguish between these possibilities we
measure the order parameter of the nematic phase in coexistence
with the isotropic phase as is shown in Fig. \ref{SCphi}. In Fig.
\ref{SCphi}a we compare the order parameter of rods with and
without attractive interactions (ie. with and without the addition
of polymer) and find that the order parameter is determined by the
concentration of rods only. The nematic order parameter is plotted
as a function of polymer concentration in Fig. \ref{SCphi}b to
illustrate the independence of the coexisting nematic order
parameter with increasing attraction. This is further confirmed
with the graph in Fig. \ref{9.fig}, which shows that the nematic
order parameter is independent of polymer concentration even well
into the nematic phase. This is in agreement with both the SVTA
and SPTA theories which predict that the measured order parameter
of the nematic rod/polymer mixture will depend only on the
concentration of rods and be independent of the level of
attraction (ie. polymer concentration). We note that the order
parameter data is noisy because of the intrinsic high viscosity of
the fd/Dextran solutions. This high viscosity makes it difficult
to create nematic monodomains even in magnetic fields up to 8T.

\section{Conclusions}

We have presented quantitative measurements of the
isotropic-nematic phase transition in a binary suspension of
rod-like particles (fd) and spherical polymers (Dextran). The
widening of the coexistence concentrations and partitioning of the
polymer predicted theoretically are observed in these experiments
on {\it fd}-Dextran mixtures. As discussed in the previous
paragraph, our measurements indicate that the liquid of rods can
be thought of as a van der Waals liquid where the order parameter
of the nematic phase is determined by repulsive interactions,
while attractive interactions provide structureless cohesive
energy. Within the admittedly noisy experimental data, we find
that the order parameter is determined solely by the rod
concentration and not by the polymer concentration, or
equivalently, the strength of attraction. However, even after
taking the following effects into account; the possibility of the
virus and polymer interpenetrating, the charge of the virus, and
the semi-flexibility of the virus we found large quantitative
differences between the theory and the experiment~\cite{Sear97}.
Notably, the theory severely overestimated the strength of the
polymer induced attraction. The difference is especially
pronounced in the nematic phase and at low ionic strength. This
and previous work~\cite{Adams98,Dogic01} suggest that much remains
to be done before we are able to understand and predict the
behavior of rods whose interactions are more complex then simple
Onsager-like hard rods.

We acknowledge the support of the National Science Foundation
NSF-DMR 0088008. We thank Pavlik Lettinga for reading of the
manuscript. Additional information, movies, and photographs are
available online at \texttt{www.elsie.brandeis.edu}.

\section{Appendix I}

In this appendix we present the calculation of the phase diagram
for a rod-polymer mixture using the SPTA
theory~\cite{Lekkerkerker94,Lekkerkerker92,Meijer91}. Several
misprints in the original paper are corrected
here~\cite{Lekkerkerker94}. The approximate free energy of the
colloid-polymer mixture is given by the following
expression~\cite{Lekkerkerker92a}:

\begin{equation}
\label{thermo_perturbation}
F_{C+P}(\phi)=F_C(\phi)-\Pi_p<V_{\mbox{\scriptsize{free}}}(\phi)>
\end{equation}

\noindent where $F_C(\phi)$ is the free energy of colloid
suspension at volume fraction $\phi$. The co-existence
concentrations for the I-N transition predicted by the scaled
particle theory are in very close agreement with the results from
the computer simulations~\cite{Kramer98}. This indicates that the
scaled particles theory provides a good approximation for third
and higher virial coefficients. The system is assumed to be in
equilibrium with a polymer reservoir which is separated from the
colloid-polymer mixture by a membrane permeable to polymers only.
The osmotic pressure of the polymers in the reservoir is
$\Pi_p=\rho k_B T$, with $\rho$ the polymer number density.
$V_{\mbox{\scriptsize free}}$ is the free volume available to a
polymer in a solution of pure hard particle colloids. It is
assumed that $V_{\mbox{\scriptsize free}}$ in a polymer/colloid
mixture is equal to the $V_{\mbox{\scriptsize free}}$ in the pure
colloid suspension. In this sense Eq.~\ref{thermo_perturbation} is
a thermodynamic perturbation theory.

The expression for the free energy of a pure hard spherocylinder
colloidal suspension is given by the scaled particle theory
developed by Cotter~\cite{Cotter79} :

\begin{eqnarray}
\label{scaled_free_energy}
    \frac{F_C(\delta,\phi,\alpha)}{Nk_bT}=&&\ln(\phi)+\ln(1-\phi)+ \sigma(\alpha,L/P)+\Pi_2(\delta,\alpha)
\frac{\phi}{1-\phi}\nonumber \\
&&+\frac{1}{2} \Pi_3(\delta,\alpha)
\left(\frac{\phi}{1-\phi}\right)^2
\end{eqnarray}

\noindent where $\phi$ is the volume fraction of spherocylinders
\begin{equation}
    \phi=\frac{N_{\mbox{\scriptsize{rods}}}}{V}
    \left(\frac{\pi}{6}D^3+\frac{\pi}{4}D^2L
    \right).
\end{equation}

\noindent The coefficients $\Pi_2$ and $\Pi_3$ are given by the
following expressions

\begin{equation}
\Pi_2(\delta,\alpha)=3+
\frac{3(\delta-1)^2}{(3\delta-1)}\xi(\alpha),
\end{equation}
\begin{equation}
\Pi_3(\delta,\alpha)=\frac{12\delta(2\delta-1)}{(3\delta-1)^2}+\frac{12\delta(\delta-1)^2}{(3\delta-1)^2}\xi(\alpha)
\end{equation}

\noindent and the parameter $\delta$ is the overall length to
diameter ratio of the spherocylinder $\delta=\frac{L+D}{D}$. The
function $\sigma(\alpha,L/P)$ is an expression that accounts for
the rotational entropy of the rods and the entropy associated with
the loss of configurations due to confinement of the bending modes
of the semi-flexible rods in the nematic phase has been derived by
extrapolating between the hard rod and the flexible chain
limits~\cite{Hentschke90,Odijk86,DuPre91}. In this paper the
expression for $\sigma$ obtained by Hentschke is used for
numerical calculations and is given by

\begin{equation}
\label{sigma}
\sigma(\alpha,\frac{L}{P})=\ln(\alpha) -1 +\pi
e^{-\alpha}+\frac{L}{6P}(\alpha-1)+\frac{5}{12}\ln\left(\cosh\left(\frac{L}{P}\frac{\alpha-1}{5}\right)\right)
\end{equation}

The function $\xi(\alpha)$ that describes the interactions between
rods at the level of second virial coefficient is given by:

\begin{eqnarray}
\label{Onsager_F}
 \xi (\alpha) &= & \frac{2I_2(\alpha)}{\sinh^2(\alpha)}
\end{eqnarray}

For this calculation we assume the Onsager ansatz for the
orientational distribution function given by :
\begin{eqnarray}
\label{Distribution}
 f(\alpha,\cos(\theta))=\frac{\alpha \mbox{cosh}(\alpha
\cos(\theta))}{4 \pi \mbox{sinh}(\alpha)}.
\end{eqnarray}

The expression for the free volume in a spherocylinder suspension
is given by:

\begin{equation}
\nu(\phi,\delta,q)=\frac{V_{\mbox{\scriptsize
free}}}{V}=(1-\phi)\mbox{exp}
\left({-\left(A(\delta,q)\left(\frac{\phi}{1-\phi}\right)+B(\delta,q)
\left(\frac{\phi}{1-\phi}\right)^2+C\left(\frac{\phi}{1-\phi}
\right)^3 \right)} \right)
\end{equation}
\noindent where
\begin{eqnarray}
A(\delta,q) & = & \frac{6\delta}{3\delta-1}+\frac{3(\delta+1)}
{3\delta-1}q^2+\frac{2}{3\delta-1}q^3,\nonumber
\\
B(\delta,q) &= &\frac{1}{2}\left(\frac{6\delta}
{3\delta-1}\right)^2q^2+\left(\frac{6}{3\delta-1}+\frac{6(\delta-1)^2}
{(3\delta-1)^2}\xi(\alpha)\right)q^3, \nonumber \\
C(\delta,q) &=&
\frac{24\delta}{3\delta-1}\left(\frac{2\delta-1}{(3\delta-1)^2}+\frac{(\delta-1)^2}
{(3\delta-1)^2}\xi(\alpha)\right)q^3.
\end{eqnarray}

\noindent The ratio of the polymer diameter to the rod diameter is
given by the parameter $q$. After the expression for the scaled
particle free energy (~\ref{scaled_free_energy}) is obtained, we
insert the Onsager approximation for the orientational
distribution functions $f(\alpha)$ and minimize the free energy at
different rod concentrations with respect to the parameter
$\alpha$ to find the order parameter of the nematic phase at that
concentration. To find the concentrations of rods in the
coexisting isotropic and nematic phases we solve the conditions
for the equality of the osmotic pressure and chemical potential.
The expressions for the osmotic pressure and the chemical
potential are :

\begin{eqnarray}
\label{coexistence}
\Pi &= &\phi^2 \frac{\partial F_c(\phi)}{\partial \phi}+ n_p \lambda \left(\nu-\phi \frac{\partial \nu({\phi})}{\partial \phi}\right) \nonumber \\
\mu  &= & F_c(\phi)+ \phi \frac{\partial F_c(\phi)}{\partial \phi}
+ n_p \lambda \frac{\partial \nu(\phi)}{\partial \phi}
\end{eqnarray}

\noindent where $n_p$ is the polymer volume fraction, $\frac{4}{3}
\pi R_g^3\rho$, and $\lambda$ is the ratio of spherocylinder
volume to polymer volume
\begin{equation}
\lambda=\frac{1}{q^3}\left(1+\frac{3}{2}(\delta-1)\right)
\end{equation}

\noindent The phase diagram is calculated by first minimizing the
free energy with the respect to the parameter $\alpha$ and then
solving coexistence equations(~\ref{coexistence}). The SVTA phase
diagrams for rigid rods are calculated following the calculation
of Warren \cite{Warren94}. To extend this calculation to
semi-flexible rods the orientational entropy term in the Onsager
free energy was replaced by the confinement entropy of
semi-flexible polymers as shown in Eq.~\ref{sigma}. To account for
electrostatics the rod diameter $D$ is replaced with
$D_{\mbox{\scriptsize eff}}$\cite{Tang95}. To correct the AO
theory for the case of a polymer radius larger than the colloid
radius we replace the polymer density $\rho$ with
$\rho_{\mbox{\scriptsize eff}}$ and the polymer radius $R_g$ with
$R^{\mbox{\scriptsize AO}}_{\mbox{\scriptsize eff}}$ as described
in the text and in Figures 3,5 and 6.

\newpage

\end{document}